\documentclass[Crown,sageh,times,doublespace]{sagej}

\usepackage{moreverb,url}

\usepackage[colorlinks,bookmarksopen,bookmarksnumbered,citecolor=red,urlcolor=red]{hyperref}
\usepackage{graphicx}
\newcommand\BibTeX{{\rmfamily B\kern-.05em \textsc{i\kern-.025em b}\kern-.08em
T\kern-.1667em\lower.7ex\hbox{E}\kern-.125emX}}

\def\volumeyear{2022}
\newtheorem{theorem}{Theorem}[section]
\begin{document}

\runninghead{Semi-parametric GEE for cross-over designs}

\title{Semi-parametric generalized estimating equations for repeated measurements in cross-over designs}

\author{N. A. Cruz\affilnum{1}, 
        O.O. Melo\affilnum{2}, 
        and C.A. Martinez\affilnum{3}}

\affiliation{\affilnum{1}PhD Student, Department of  Statistics,
      Faculty of Sciences,
      Universidad Nacional de Colombia,
      neacruzgu@unal.edu.co\\
\affilnum{2}PhD,
      Associate Professor,
      Department of  Statistics,
      Faculty of Sciences, Universidad Nacional de Colombia, oomelom@unal.edu.co
\affilnum{3}PhD,Associate Researcher,
      Corporación Colombiana de Investigación Agropecuaria – AGROSAVIA, Sede Central, cmartinez@agrosavia.co}

\corrauth{N.A. Cruz, 
            Corresponding author,
      PhD Student, Department of  Statistics,
      Faculty of Sciences,
      Universidad Nacional de Colombia,
      Phone:(+571)\;3165000 Ext: 13206,
      (+571)\;3165000\; Ext: 13210}

\email{neacruzgu@unal.edu.co}

\begin{abstract}
%A model for cross-over designs with repeated measures within each period was developed. It is obtained using an extension of generalized estimating equations that includes a parametric component to model treatment effects and a non-parametric component to model time and carry-over effects, the estimation approach for the non-parametric  component is based on splines. A simulation study was carried out to explore the model properties.  Thus, when there isn a carry-over effect or a functional temporal effect, the proposed model presents better results than the standard models.  The model estimates significant carry-over effects in the blood pressure data set, and allows a better estimation of the effects of treatment and period that the standard models, additionally it is applied in an insulin experiment in rabbits. Among the theoretical properties, the solution is found to be analogous to weighted least squares. Therefore, model diagnostics can be made adapting the results from a multiple regression.
A model for cross-over designs with repeated measures within each period was developed. It is obtained using an extension of generalized estimating equations that includes a parametric component to model treatment effects and a non-parametric component to model time and carry-over effects; the estimation approach for the non-parametric  component is based on splines. A simulation study was carried out to explore the model properties.  Thus, when there is a carry-over effect or a functional temporal effect, the proposed model presents better results than the standard models. Among the theoretical properties, the solution is found to be analogous to weighted least squares. Therefore, model diagnostics can be made adapting the results from a multiple regression. 
%The proposed methodology was implemented in two real datasets: one on systolic blood pressure data, and the other, on insulin in rabbits.
The proposed methodology was implemented in the data sets of the crossover experiments that motivated the approach of this work: systolic blood pressure  and  insulin in rabbits.
% La metodología propuesta fue implementada en los conjuntos de datos de los experimentros crossver que motivaron el planteamiento de esta metodología.  
\end{abstract}


\maketitle

\section{Introduction}
%En el contexto de los diseños experimentales crossover,  cada unidad experimental recibe una secuencia de tratamientos, y cada tratamiento es aplicado durante un periodo de tiempo. Estos diseños son muy útiles en la experimentación médica, pues requieren de menos unidades experimentales para obtener los mismos resultados que estudios transversales. La desventaja se da por la aparición de efectos carry-over, que se definen cómo los efectos residuales que quedan en la respuesta del individuo y que son causados por los tratamientos aplicados en los periodos anteriores (\cite{madeyski2018effect} and \cite{kitchenham2018corrections}).    

In the context of crossover experimental designs, each experimental unit receives a sequence of treatments, and each treatment is applied over a period of time \citep{biabani2018crossover}. These designs are very useful in medical experimentation, since they require fewer experimental units to obtain the same results as cross-sectional studies. The disadvantage is given by the appearance of carry-over effects, which are defined as the residual effects that remain in the response of the individual and that are caused by the treatments applied in the previous periods (\cite{madeyski2018effect} and \cite{kitchenham2018corrections}).

%Los trabajos más recientes sobre el análisis de diseños crossover asumen la inexistencia de los efectos carry-over, debido a la  presencia de un periodo de lavado entre las aplicaciones sucesivas de tratamientos  \cite{curtin2017meta}. Este supuesto es común en los trabajos basados en modelos lineales generalizados clásicos  \cite{li2018power},  en modelos bayesianos \cite{oh2003bayesian} o en modelos de ecuaciones de estimación generalizadas (GEE) \cite{curtin2017meta}.  Sin embargo, en algunos diseños crossover la longitud del periodo de lavado es muy corta y no garantiza la eliminación de los efectos residuales de cada uno de los tratamientos, cómo el presentado en  \citet[pag 204]{ken15}. En este diseño 3 tratamientos para el control de la presión arterial son usados y  el tratamiento C es un placebo, por lo tanto, si hay efecto carry-over del placebo, este no se limpia en el periodo de lavado. Además, en el tratamiento de la hipertensión, los tratamientos se pueden suspender por muy poco tiempo debido a las características de la enfermedad.  
The most recent works on the analysis of crossover designs assume the non-existence of carry-over effects, due to the presence of a washout period between the successive applications of treatments \citep{curtin2017meta}. This assumption is common in works based on classical generalized linear models \citep{li2018power}, Bayesian models \citep{oh2003bayesian} or generalized estimating equation (GEE) models \citep{curtin2017meta}. However, in some crossover designs the length of the washout period is very short and does not guarantee the elimination of the residual effects of each of the treatments, such as the one presented in \citet[page 204]{ken15}. In this design, three treatments for blood pressure control are used and treatment C is a placebo. In this experiment, if there is a carry-over effect of the placebo, it is not cleaned in the washout period. The doctors tried to control the hypertension in patients, and so, treatments can be stopped for a very short time due to the characteristics of the disease.

 %% treat to the control the hypertension, and the can not be stopped for a very short time
%Adicionalmente, en el diseño presentado en \citet[pag 204]{ken15} la presión arterial sistólica es observada 10 veces dentro de cada periodo:  30 y 15 minutos antes de la aplicación, y 15, 30, 45, 60, 75, 90, 120 y 240 minutos después de la aplicación, como se ve en la tabla \ref{tabla100}, lo que genera una medida repetida por cada periodo de aplicación. Este tipo de diseños se conoce como diseño crossover con medidas repetidas \citep{dubois2011model}. \cite{dubois2011model}, \cite{diaz2013random} y \cite{for15} used Gaussian linear mixed models to study cross-over repeated measures designs. However, those studies considered one observation per period by calculating the area under the curve, and they did not include the carry-over effects. Sin embargo, este modelamiento no permite observar el comportamiento temporal de la variable respuesta dentro del periodo, ni tampoco la presencia de efectos carry-over que fluctuen en el tiempo. 
Furthermore, in the design presented in \citet[pag 204]{ken15}, the systolic blood pressure is observed ten times within each period: 30 and 15 minutes before the application, and 15, 30, 45, 60, 75, 90 , 120 and 240 minutes after the application, as seen in the Table \ref{tabla100}, which generates a repeated measurement for each application period. This type of design is known as a repeated measures crossover design \citep{dubois2011model}. \cite{dubois2011model}, \cite{diaz2013random} and \cite{for15} used Gaussian linear mixed models to study cross-over repeated measures designs. However, those studies considered one observation per period by calculating the area under the curve, and they did not include the carry-over effects. Additionally, this modeling does not allow us to observe the temporal behavior of the response variable within the period, nor the presence of carry-over effects that fluctuate over time.
%Por último, cuando se analiza la variable respuesta del experimento crossover mostrado por \cite{10.1093/biostatistics/kxp046}  se observa que no se ajusta de manera adecuada a una distribución normal.  Debido a que la respuesta de este experimento son los niveles de azucar en sangre, los cuales presentan asimetria y son siempre positivos, unda distribucuión gamma parece más adecuada para su analisis. 

On the other hand, when the response variable of the crossover experiment shown by \cite{10.1093/biostatistics/kxp046} is analyzed, it is observed that it does not adequately fit a normal distribution. Since the response in this experiment is blood sugar levels, which are skewed and always positive, a gamma distribution seems more suitable for analysis. In both experiments, we have a response that can be assumed to be in the exponential family and the responses of the same experimental unit are correlated. Therefore, in this paper we propose an extension of the GEE with splines to model the effects of main interest in the design (treatments, period) with a parametric component and the temporary effects through smoothing splines.
%Esta metodología permite aislar de manera insesgada el comportamiento temporal del efecto carry-over de los efectos de periodo y de tratamiento, lo cuál se demuestra de manera teórica y mediante un ejercicio de simulación. Posteriormente al aplicar la metodología en el diseño de presión arterial, se obtiene un efecto carry-over significativo del tratamiento placebo, corroborando la importancia de tenerlo en cuenta en el análisis.

This methodology makes it possible to unbiasedly isolate the temporal behavior of the carry-over effect from the period and treatment effects, which is demonstrated theoretically and through a simulation exercise. Subsequently, when applying the methodology in the blood pressure design, a significant carry-over effect of the placebo treatment is obtained, corroborating the importance of taking it into account in the analysis.
%In the context of crossover experimental designs, each experimental unit receives a sequence of treatments, and each treatment is applied over a period of time \citep{biabani2018crossover}. These are very useful in medical experimentation, since they require fewer experimental units to obtain the same results as cross-sectional studies, although with the disadvantage of the appearance of carry-over effects. Most of the works that address the analysis of these designs assume that there is a washout period between the successive applications of treatments and, furthermore, each experimental unit is observed once in each period. These two assumptions allow modeling the response without including carry-over effects, since it is assumed that the carry-over effect is eliminated in the washout period  (a period of time between treatments with no treatment at all)  (\cite{madeyski2018effect} and \cite{kitchenham2018corrections}). 

%However, 
\begin{table}[ht]
    \centering
    \begin{tabular}{c|c|c|c|}
        Sequence & Period 1 & Period 2 & Period 3\\
        \hline
        \begin{tabular}{cc}
            (1) ABC & Ind 1\\
             & Ind 2
        \end{tabular} & \begin{tabular}{c}
               10 observations\\
              10 observations
        \end{tabular}&  \begin{tabular}{c}
              10 observations\\
              10 observations
        \end{tabular}&  \begin{tabular}{c}
              10 observations\\
             10 observations
        \end{tabular}\\
        \vdots & \vdots &\vdots &\vdots \\
         \begin{tabular}{cc}
            (6) CBA & Ind 11\\
             & Ind 12
        \end{tabular} & \begin{tabular}{c}
               10 observations\\
              10 observations
        \end{tabular}&  \begin{tabular}{c}
              10 observations\\
              10 observations
        \end{tabular}&  \begin{tabular}{c}
              10 observations\\
             10 observations
        \end{tabular}\\
        \hline
    \end{tabular}
    \caption{Structure of the blood pressure crossover design, with three period, six sequences and ten measurements (that were taken -30, -15, 15, 30, 45, 60, 75, 90, 120 and 240 minutes from application) per period}
    \label{tabla100}
\end{table}
This paper is structured as follows: In Section 2, the semiparametric model with GEE was described, and the estimation equations are derived. In section 3, asymptotic consistency and unbiasedness of estimators are established. In Section 4, a simulation study is carried out to display the advantages of the proposed model over those models often found in the literature and some diagnostics measures for its residuals. In Section 5, an application of to blood pressure data is performed out to illustrate the model properties  and to carry out an overall analysis of this dataset. Finally, some conclusions are presented in Section 6.

\section{Repeated measures cross-over design}
A cross-over design entails five components \citep{ken15}: i) sequences which are randomly assigned combinations of the applied treatments on the experimental units, ii) treatments, that are applied to each experimental unit as a part of a sequence in a given time, iii) periods, that represent the application lapse for the treatments which are part of a sequence,  v) experimental units, which are the elements on which a treatment is applied.\\
In each sequence, there are $n_l$ experimental units, therefore the total number of experimental units is $n=\sum_{l=1}^S n_l$ Further, it is frequent that each period has the same length for all sequences, therefore, the number of observation periods equals the length of each sequence.\\
 For the structure of cross-over designs, the carry-over effects constitute part of them. \cite{vegas2016crossover} defined the carry-over as a treatment’s effect persistence over those treatments applied later.  That is, if a treatment is applied on a given period, then there exists the possibility of a residual or carry-over effect that persists in the following periods when other treatments are applied. When the carry-over effect of a treatment affects the one applied in the next period, it is known as a frist-order carry over effect. \\
In a cross-over design with $S$ sequences of length $P$, let $n_{ij}$ be the number of observations on the $i$-th experimental unit and $j$-th period, then  $\pmb{Y}_{ij}$ is a vector defined as:
\begin{equation}\label{observacion}
  \pmb{Y}_{ij}=\left(Y_{ij1}, \ldots, Y_{ijn_{ij}} \right)^T
 \end{equation}
Moreover, we define a vector $\pmb{Y}_{i}$ vector contains every observation on the $i$-th experimental unit
 \begin{equation}
\pmb{Y}_{i}= \left(\pmb{Y}_{i1}, \ldots, \pmb{Y}_{iP}\right)^T
 \end{equation}
and its size is $\sum_{j=1}^P n_{ij}$.

Regarding the use of smoothing functions, \citet{wild1996additive} proposed a kernel smoothing to select explanatory variables in GEE models, \citet{lin2001semiparametric} derived a semiparametric estimation equation for repeated measures data and presented some asymptotic properties without including the correlation matrix.  On the other hand,  \citet{he2002estimation}  presented a semiparametric model with correlated normal data and explored the properties of symmetric kernels, \citet{stoklosa2018generalized} developed a GEE generalization for adaptive multivariate splines through m-estimators, and \cite{yang2021semi} discussed a GEE model with two semiparametric functions for normally distributed responses and some kernel smoothing functions. 
 
 Accordingly, GEE will be used because $Y_{ijk}$ (the response variable) has a distribution that belongs to the exponential family, % which can be normal or Poisson distributions 
 and also a semiparametric model with B-splines for the time and carry-over effects as follows:
 \begin{align}
     E(Y_{ijk})&=\mu_{ijk}, \qquad Var(Y_{ijk})=\phi V(\mu_{ijk})\nonumber \\
     g(\mu_{ijk})&=\pmb{x}_{ijk}^T\pmb{\beta} + f_1(\pmb{Z}_{1ijk})+ f_2(\pmb{Z}_{2ijk}) \label{ecverdad}\\
     \pmb{V}(\pmb{\mu}_i)&=\left[\pmb{D}(V(\mu_{ijk})^{\frac{1}{2}}) \pmb{R}(\pmb{\pmb{ \alpha}}) \pmb{D}({V}(\mu_{ijk})^{\frac{1}{2}})\right]_{P \times P} \label{ecverdad1}
\end{align}
where $g(\cdot)$ is the link function associated to the exponential family, $\pmb{x}_{ijk}$  is  the vector of the design matrix associated to the $k$-th response of the $i$-th  experimental unit in the $j$-th  period, $\pmb{\beta}$ represents the parametric effects,  $f_1$  is a function describing the time’s effect period, $f_2$ is a function describing the previous treatment carry-over effect on the current period (with $f_2(\pmb{Z}_{2i1k})=0$), $V(\mu_{ijk})$  is the variance function related to the exponential family, and $R(\pmb{ \alpha})$  is the associated correlation matrix.

Let $\{s_1(t), \ldots, s_{m}(t)\}$ be a basis splines, then  the  $f_1$ and $f_2$ functions can be approximated through the following equations \citep{yu2012spline}
\begin{align}
    \hat{f}_1(t)&= \sum_{b=1}^m \hat{\alpha}_{1b}s_b(t) \label{spli1}\\
    \hat{f}_2(t)&= \sum_{b=1}^m \hat{\alpha}_{2b}s_b(t)\label{spli2}
\end{align}
where $m=max(n_{ij})$. Adapting the estimation equations given by \citet{he2002estimation}, the following generalized estimation equations are proposed for $\alpha_{1}=\{\alpha_{11},\ldots, \alpha_{1m}\}$, $ \alpha_2=\{\alpha_{21},\ldots, \alpha_{2m}\}$  and  $\pmb{\beta}$:
\begin{itemize}
    \item For the time effect 
    \begin{align}
    &U_1(\pmb{\alpha_1}, t|\pmb{\beta}, \pmb{\alpha}_2, \pmb{\alpha})=\nonumber\\
    &\sum_{i=1}^n \left\{ diag\left(\frac{\partial \mu_{ijk}}{\partial \pmb{\alpha}_1}\right)\right\}_i \frac {\pmb{V}^{-1}_{1i}}{\phi} \left(\pmb{y}_i-\pmb{\mu}_i\left[\pmb{X}_i \pmb{\beta},\sum_{b=1}^m \alpha_{1b}s_b(t)  ,\hat{f}_2(\pmb{Z}_{2i}) \right]\right) \label{u1}
\end{align}
where
\begin{align}
    &\pmb{V}_{1i}=\left\{diag\left(V \left(\pmb{\mu}_i\left[\pmb{X}_i \pmb{\beta},\sum_{b=1}^m \alpha_{1b}s_b(t)  ,\hat{f}_2(\pmb{Z}_{2i}) \right]  \right) \right)\right\}_i^\frac{1}{2} \times \pmb{R}(\mathbf{\pmb{ \alpha}})\times\nonumber\\ 
    & \left\{diag\left(V \left(\pmb{\mu}_i\left[\pmb{X}_i \pmb{\beta},\sum_{b=1}^m \alpha_{1b}s_b(t)  ,\hat{f}_2(\pmb{Z}_{2i}) \right]  \right) \right)\right\}_i^\frac{1}{2}, \qquad   i=1, \ldots, n  \nonumber
\end{align}
$\left\{ diag\left(\frac{\partial \mu_{ijk}}{\partial \pmb{\alpha}_1}\right)\right\}_i$ is a diagonal matrix with elements on the diagonal given by 
\begin{equation}\nonumber
    \left\{\frac{\partial \mu_{i11}}{\partial \pmb{\alpha}_1},\frac{\partial \mu_{i12}}{\partial \pmb{\alpha}_1}, \ldots, \frac{\partial \mu_{i1n_{i1}}}{\partial \pmb{\alpha}_1}, \ldots, \frac{\partial \mu_{iPn_{iP}}}{\partial \pmb{\alpha}_1} \right\}
\end{equation}
and $V(\cdot)$ is the variance function of the exponential family applied to each of the $i$-th individual's expected values.
\item For the carry-over effects
\begin{align}
   &U_2(\pmb{\alpha_2}, t|\pmb{\beta}, \pmb{\alpha}_1, \pmb{\alpha})=\nonumber\\
  & \sum_{i=1}^n \left\{ diag\left(\frac{\partial \mu_{ijk}}{\partial \pmb{\alpha}_2}\right)\right\}_i\frac {\pmb{V}^{-1}_{2i}}{\phi} \left(\pmb{y}_i-\pmb{\mu}_i\left(\pmb{X}_i \pmb{\beta},\sum_{b=1}^m \alpha_{2b}s_b(t)  ,\hat{f}_1(\pmb{Z}_{1i}) \right)\right) \label{u2}
  \end{align}
where
\begin{align}
    &\pmb{V}_{2i}=\left\{diag\left(V \left(\pmb{\mu}_i\left[\pmb{X}_i \pmb{\beta},\sum_{b=1}^m \alpha_{2b}s_b(t)  ,\hat{f}_1(\pmb{Z}_{1i}) \right]  \right) \right)\right\}_i^\frac{1}{2} \times \pmb{R}(\mathbf{\pmb{ \alpha}})\times\nonumber\\ 
    & \left\{diag\left(V \left(\pmb{\mu}_i\left[\pmb{X}_i \pmb{\beta},\sum_{b=1}^m \alpha_{2b}s_b(t)  ,\hat{f}_1(\pmb{Z}_{1i}) \right]  \right) \right)\right\}_i^\frac{1}{2}, \qquad   i=1, \ldots, n   \nonumber
\end{align}
\item For the fixed effects, that is, treatment, sequence, period or other covariates
\begin{align}
    &U_3(\pmb{\beta}|\pmb{\alpha}_1, \pmb{\alpha}_2, \pmb{\alpha})=\nonumber\\
    &\sum_{i=1}^n \left\{  diag\left(\frac{\partial \mu_{ijk}}{\partial \pmb{\beta}}\right)\right\}_i\frac {\pmb{V}^{-1}_{3i}}{\phi} \left(\pmb{y}_i-\pmb{\mu}_i\left[\pmb{X}_i \pmb{\beta},\hat{f}_1(\pmb{Z}_{1i} ,\hat{f}_2(\pmb{Z}_{2i}) \right]\right) \label{u3}
    \end{align}
where
\begin{align}
    &\pmb{V}_{3i}=\left\{diag\left(V \left(\pmb{\mu}_i\left[\pmb{X}_i \pmb{\beta},\hat{f}_1(\pmb{Z}_{1i})   ,\hat{f}_2(\pmb{Z}_{2i}) \right]  \right) \right)\right\}_i^\frac{1}{2}\times \pmb{R}(\mathbf{\pmb{ \alpha}})\times\nonumber\\ & \left\{diag\left(V \left(\pmb{\mu}_i\left[\pmb{X}_i \pmb{\beta},\hat{f}_1(\pmb{Z}_{1i})   ,\hat{f}_2(\pmb{Z}_{2i}) \right]  \right) \right)\right\}_i^\frac{1}{2}, \qquad   i=1, \ldots, n  \nonumber
\end{align}
\item For the correlation matrix 
\begin{equation}\label{u4}
 U_4(\pmb{ \alpha}|\pmb{\beta}, \pmb{\alpha}_1, \pmb{\alpha}_2)=\sum_{i=1}^S \sum_{k=1}^{n_i}\left(\frac{\partial \pmb{\varepsilon}_{ik}}{\partial \pmb{ \alpha}} \right)^T \pmb{F}_{ik}^{-1} \left(\pmb{W}_{ik} - \pmb{\varepsilon}_{ik} \right)
\end{equation}
where $\pmb{F}_{ik}=\pmb{D}(V(r_{ijk}))_{q\times q}$ is a diagonal matrix, $\pmb{\varepsilon}_{ik}=E(\pmb{W}_{ik})_{q\times 1}$ and 
$\pmb{W}_{ik}= (r_{i1k}r_{i2k},$ $r_{i1k}r_{i3k},\ldots,r_{i(T-1)k}r_{iTk})^T_{q\times 1}$,  $r_{ijk}$ is the $ijk$-th  Pearson residual and  $q={T \choose 2}$.
\end{itemize}
To get the estimators of $\pmb{\alpha}_1$, $\pmb{\alpha}_2$, $\pmb{\beta}$ and $\pmb{\alpha}$ the following steps are performed:
\begin{enumerate}
    \item Set initial values  $\pmb{\alpha}_{2}^{(0)}$, $\pmb{\beta}^{(0)}$ y $\pmb{\alpha}^{(0)}$
    \item Find the value $\pmb{\alpha}_{1}^{(1)}$ that solves the equation
    \[U_1(\pmb{\alpha_1}, t|\pmb{\beta}^{(0)}, \pmb{\alpha}_2^{(0)}, \pmb{\alpha}^{(0)})=0\]  
 \item Find the value $\pmb{\alpha}_{2}^{(1)}$ that solves the equation
 \[U_2(\pmb{\alpha_2}, t|\pmb{\beta}^{(0)}, \pmb{\alpha}_1^{(1)}, \pmb{\alpha}^{(0)})=0\]  
 \item Find the value $\pmb{\beta}^{(1)}$ that solves the equation
 \[U_3(\pmb{\beta}|\pmb{\alpha}_1^{(1)}, \pmb{\alpha}_2^{(1)}, \pmb{\alpha}^{(0)})=0\]  
  \item Find the value $\pmb{\alpha}^{(1)}$ that solves the equation
 \[U_4(\pmb{\alpha}|\pmb{\beta}^{(1)},\pmb{\alpha}_1^{(1)}, \pmb{\alpha}_2^{(1)}, \pmb{\alpha}^{(0)})=0\]
 \item Repeat steps (2) to (5) until convergence.
\end{enumerate}
To solve  equation \eqref{u3}, the Fisher scoring algorithm is used, that is, in the  $m$-th step, the estimator of $\pmb{\beta}$ is given by:
\begin{align}
 \pmb{\beta}^{(m+1)}&=\pmb{\beta}^{(m)}- \left[E\left\{ U_3'(\pmb{\beta}^{(m)}|\pmb{\alpha}_1, \pmb{\alpha}_2, \pmb{\alpha})\right\} \right] ^{-1} U_3(\pmb{\beta}^{(m)}|\pmb{\alpha}_1, \pmb{\alpha}_2, \pmb{\alpha})\nonumber\\
 &=\pmb{\beta}^{(m)}- \left\{ \sum_{i=1}^n \pmb{X}_i^T \pmb{W}_i^{(m)} \pmb{X}_i \right\}^{-1} \left\{\sum_{i=1}^n\pmb{X}_i^T \pmb{W}_i^{(m)} \left(\pmb{N}_i^{(m)} \right)^{-1}  \pmb{u}_i^{(m)}  \right\} \label{ecmcp}
 \end{align}
 where
 \begin{align*}
    E\left\{U_3'(\pmb{\beta}^{(m)}|\pmb{\alpha}_1, \pmb{\alpha}_2, \pmb{\alpha})\right\}&= E\left\{\frac{\partial U_3 (\pmb{\beta}^{(m)}|\pmb{\alpha}_1, \pmb{\alpha}_2, \pmb{\alpha})}{\partial \pmb{\beta}}\right\}=\sum_{i=1}^n\pmb{X}_i^T \pmb{W}_i^{(m)} \pmb{X}_i,\\
\pmb{W}_i&=\left\{  diag\left(\frac{\partial \mu_{ijk}}{\partial \pmb{\beta}}\right)\right\}_i\pmb{V}^{-1}_{3i} \left\{  diag\left(\frac{\partial \mu_{ijk}}{\partial \pmb{\beta}}\right)\right\}_i ^{-1},\\
\pmb{N}_i&=\left\{  diag\left(\frac{\partial \mu_{ijk}}{\partial \pmb{\beta}}\right)\right\}_i,\\
 \pmb{u}_i&=\left(\pmb{y}_i-\pmb{\mu}_i\left[\pmb{X}_i \pmb{\beta},\hat{f}_1(\pmb{Z}_{1i} ,\hat{f}_2(\pmb{Z}_{2i}) \right]\right).
\end{align*}
Carrying out procedure similar to \cite{tsuyuguchi2020analysis}, Equation \eqref{ecmcp} can be written as:
\begin{equation}\label{betafinal}
    \pmb{\beta}^{(m+1)}=\left\{ \sum_{i=1}^n \pmb{X}_i^T \pmb{W}_i^{(m)} \pmb{X}_i \right\}^{-1} \left\{\sum_{i=1}^n\pmb{X}_i^T \pmb{W}_i^{(m)}  \pmb{z}_i^{(m)}  \right\} 
\end{equation}
where
\begin{equation}\label{eczi}
     \pmb{z}^{(m)}_i=\pmb{X}_i\pmb{\beta}^{(m)}-\pmb{N}_i^{-1(m)}\pmb{u}_i^{(m)}, \qquad   i=1, \ldots, n  
\end{equation}
Therefore, $\pmb{\hat{\beta}}$ is obtained analogously to a weighted least squares solution on the transformed response variable $\pmb{z}_i$, where the effects of the variables associated to time and the  carry-over effect have  been removed.  With these considerations, the asymptotic theory of estimators is developed using the following theorem:
\begin{theorem}\label{teorema1}
 Under the assumption that the $r$-th  derivative of $f_1$ and $f_2$ is bounded for some $r\geq 2$ and that the number of knots $m=m_n\rightarrow \infty$, but $\frac{m}{n}\rightarrow 0$ then $\hat{\pmb{\beta}}-\pmb{\beta} \xrightarrow{n\rightarrow\infty}\pmb{0}$. Also, if $m=O\left( n^{\frac{1}{(2^r+1)}}\right)$ then:
\begin{align}
   & \frac{1}{n}\sum_{i=1}^n \sum_{j=1}^{n_i} \left\{\sum_{b=1}^m \hat{\alpha}_{1b}s_b(\pmb{Z}_{1ijk})- f_1(\pmb{Z}_{1ijk})\right\}^2 = O\left(n^{-\frac{2r}{(2r+1)}}\right) \nonumber\\
   & \frac{1}{n}\sum_{i=1}^n \sum_{j=1}^{n_i} \left\{\sum_{b=1}^m \hat{\alpha}_{1b}s_b(\pmb{Z}_{2ijk})- f_2(\pmb{Z}_{2ijk})\right\}^2 = O\left(n^{-\frac{2r}{(2r+1)}}\right)\nonumber\\
  &  \sqrt{n}(\hat{\pmb{\beta}}-\pmb{\beta}) \rightarrow N(0, \pmb{A}^{-1}\pmb{B}\pmb{A}^{-1})\nonumber
\end{align}
where
\begin{align}
 \pmb{A}&=\sum_{i=1}^n  \pmb{N}_i \pmb{V}^{-1}_{3i}\pmb{N}_i^T\label{an} \\
   \pmb{B}&=\sum_{i=1}^n  \pmb{N}_i \pmb{V}^{-1}_{3i}\left(\pmb{y}_i-\pmb{\hat{\mu}}_i\right)\left(\pmb{y}_i-\pmb{\hat{\mu}}_i\right)^T \pmb{V}^{-1}_{3i} \pmb{N}_i^T \label{bn}
\end{align}
\end{theorem}
\begin{proof}
 See appendix \hyperref[Apen1]{1}
\end{proof}
\section{Model diagnostics}
\subsection{Selection}
In order to compare the fit of the proposed model model (Equations \eqref{ecverdad} and \eqref{ecverdad1}) with the fit of conventional models, then the quasi-likelihood criterion ($QIC$) defined by \citet{panq} is used:
\begin{equation}\label{QIC}
    QIC=-2QL(\hat{\pmb{\mu}};\pmb{I})+2trace(\hat{\pmb{\Omega}}^{-1}_I\hat{\pmb{V}}_{\pmb{R}} )
\end{equation}
where $\hat{\mu}=\hat{\eta}=g^{-1}(x\hat{\beta})$ is the estimated expected value for the observation with the model assuming the correlation matrix $\pmb{R}$ and the estimates are obtained using Equation \eqref{betafinal}, $\hat{\pmb{\Omega}}_I$ is the estimated variance matrix for  vector $\pmb{\beta}$ under a correlation matrix $\pmb{R}(\pmb{\alpha})=\mathbb{I}_{n_i}$, and $\hat{\pmb{V}}_{\pmb{R}}$ is the variance matrix estimated for the vector $\pmb{\beta}$ assuming the correlation matrix $\pmb{R}(\pmb{\alpha})$ as in  Equation \eqref{bn}.
  After fitting several models, the model with lowest $QIC$ is selected because, it is the one featuring the best balance between goodness of fit and complexity.
 
\subsection{Residuals}
Due to the similarity of the proposed estimator of $\pmb{\beta}$ and the weighted least squares, with weigthts given by matrix $\pmb{W}_i$, Pearson standardized residuals are proposed to assess model validity:
\begin{equation}\label{pearsonres}
    r_{ijk}=\frac{\pmb{e}_{ijk}^T\pmb{\hat{W}}_i^{\frac{1}{2}}(\pmb{\hat{z}}_i-\pmb{X}_i\pmb{\hat{\beta}})}{1-\sqrt{\hat{h}_{ijk}}}
\end{equation}
where $\pmb{e}_{ijk}$ is a vector of $n_{ij}$ zeros, except at position $k$, $\pmb{\hat{W}}_i^{\frac{1}{2}}$ is the square root of the matrix $\pmb{\hat{W}}_i$ and $h_{ijk}$ is the $ijk$ element of the diagonal of the projection matrix  $\mathcal{H}$, which is:
\begin{align}
    \mathcal{H}&=diag\left\{\pmb{H}_1, \cdots, \pmb{H}_n\right\}
    \end{align}
    with
    \begin{align}
    \pmb{H}_i&= \pmb{{W}}_i^{\frac{1}{2}}\pmb{X}_i (\pmb{X}_i^T \pmb{W}_i \pmb{X}_i)^{-1}\pmb{X}_i^T \pmb{W}_i
\end{align}
According to \citet{tsuyuguchi2020analysis}, the residuals defined in \eqref{pearsonres} are asymptotically normal with zero mean and standard deviation close to 1. Therefore, these can be used to validate the fitted  model and the conditional distribution assumption of $Y$.
\section{Simulation study}

For the simulation study, the cross-over design with extra period \citep{ken15} defined in Table \ref{tabla10} will be used, so, assume the following model:
\begin{align*}
    g(\mu_{ijk})&=\pmb{x}_{ijk}^T\pmb{\beta}+c_1 cos(t_{jk})+c_2 sen(t_{jk})\delta_{jk}
\end{align*}
where $\pmb{\beta} = (\beta_0, \beta_1, \beta_2, \beta_3)$, $\beta_0  \sim N(0,1), \; \beta_1\in (0.5,1,2)$, $\beta_2=\beta_3=3$, $ c_1 \sim N(0,1)$, $c_2 \sim N(0,1)$, $i= 1, \ldots, n$, $j=1,\ldots, 3 $ $k=1, \ldots, 3$, $n=2, \ldots, 100$ and $ Y_{ijk}\sim Poisson(\mu_{ijk})$.\\
 The number of individuals per sequence is varied from 2 to 50. $\beta_0$ is the mean, $\beta_1$ is the difference between treatments A and B, $\beta_2$ and $\beta_3$ are the effects of period 2 and 3, respectively. The  time effect is modeled by $c_1 cos(t_{jk})$ and the carry-over effect of treatment A on B is modeled by $c_2 sin(t_{jk})$. The $\delta_{jk}$ is equal to 0 in period 1 $(j=1)$ and, it is equal to 1 if in the previous period, the individual received treatment A, that is, it is a first-order carry-over effect \citep{pat51}.
\begin{table}[h!]
 \centering
 \captionsetup{justification=centering,margin=2cm}
 \begin{tabular}{|c| c c c|} 
 \hline
 Sequence & Period 1 & Period 2 & Period 3 \\ \hline
ABA & 15 observations & 15 observations & 15 observations \\
BAB & 15 observations & 15 observations & 15 observations \\
 \hline
 \end{tabular}
\caption{Cross-over design with extra period}
\label{tabla10}
\end{table}
\begin{figure}[ht]
\centering % centering figure
\includegraphics[width=12cm]{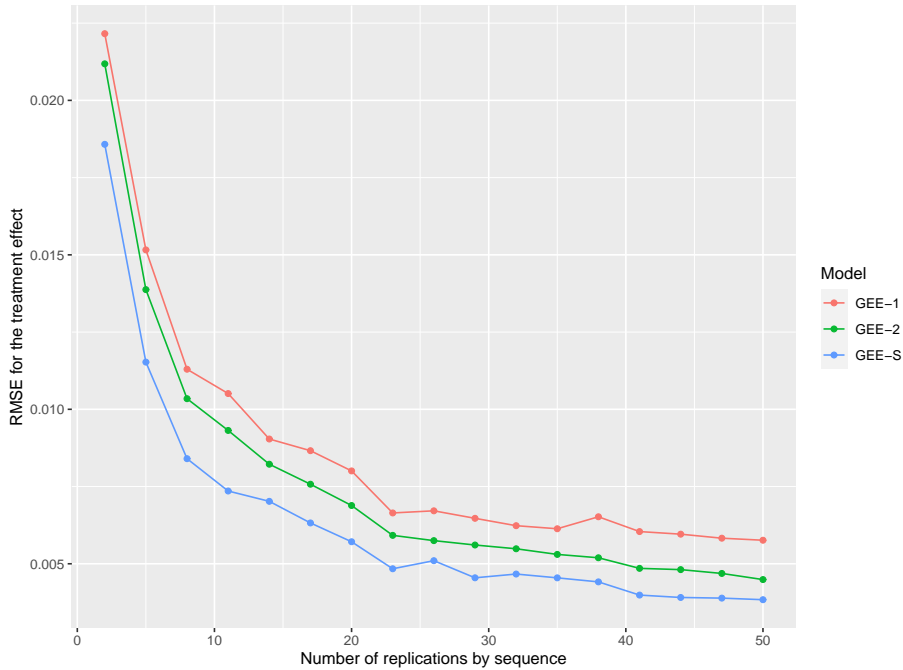}
\caption{RMSE for the estimate of $\beta_1=0.5$ in each of the models}
\label{beta1RMSE}
\end{figure}

\begin{figure}[ht]
\centering % centering figure
\includegraphics[width=12cm]{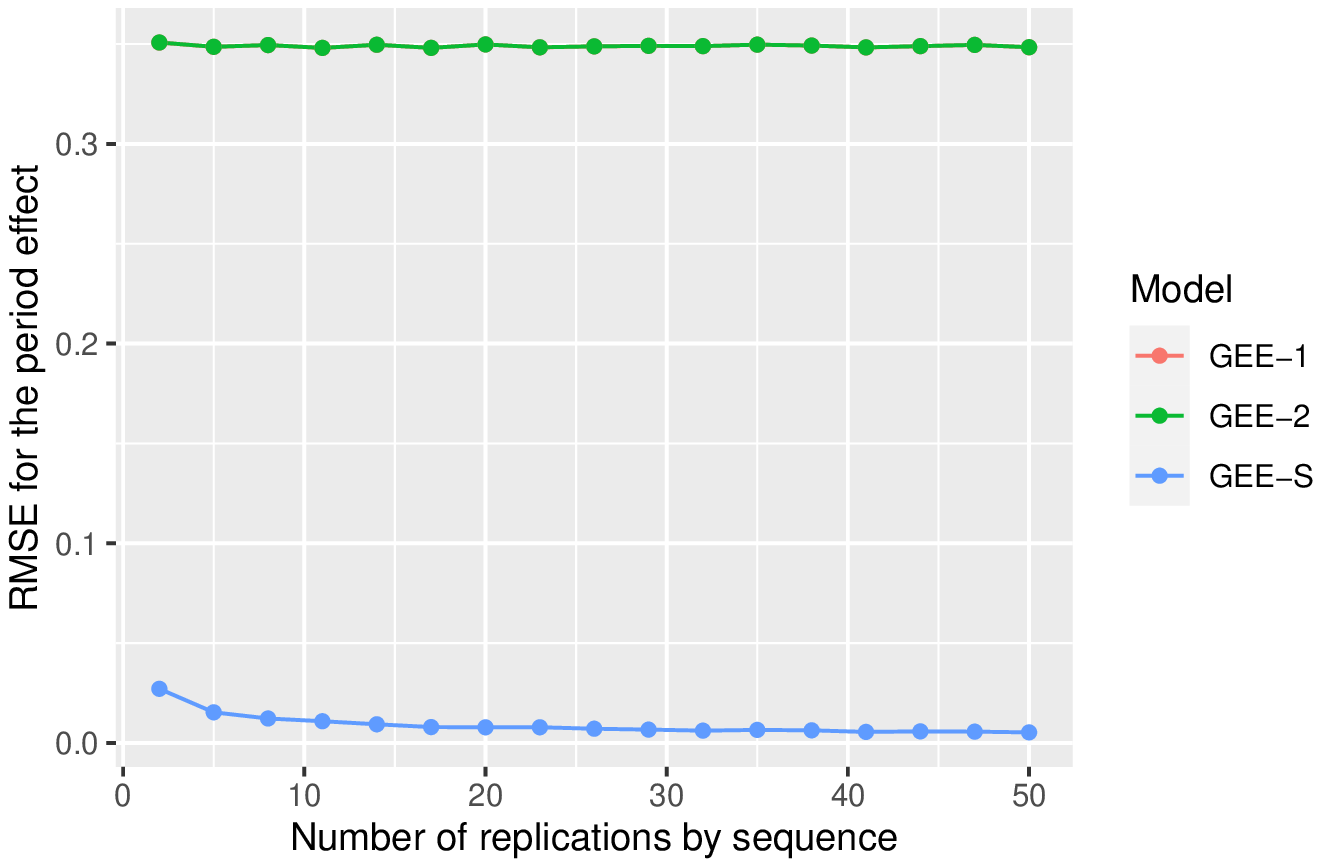}
\caption{RMSE for the estimate of $\beta_2=3$ in each of the models}
\label{beta2RMSE}
\end{figure}

Each scenario is simulated 1000 times with an autoregressive correlation matrix of order 1, and for each component of $\pmb{\beta}$, the following goodness of fit measures are obtained: 1) The root mean square error $RMSE=\sqrt{ (\hat{\beta}_i-\beta_i)^2}$ and 2) the percentage of times the hypothesis $H_0:\; \beta_1=0.5$, $H_0:\; \beta_1=1$ and $H_0:\; \beta_1=2$ are not rejected at 95\% confidence bands.

Further, for each scenario, the following three models are fitted: 1) A model defined by Equation \eqref{ecverdad} which is denoted by GEE-S, 2) a GEE model where the effect of time is linear which is denoted by GEE-1 and 3) a GEE model with quadratic time effect which is denoted by GEE-2.

Table \ref{tablebeta} shows the coverage obtained for three different values of the treatment effect ($\beta_1$) using 95\% confidence intervals. A coverage close to 95\% is observed for replicate sizes greater than 8 in the model with Splines; the other two models show coverage equal to 100\%, that is, these models over-estimate the variance of the estimated treatment effects. This is confirmed in Figure \ref{beta1RMSE}, where it is shown that although the RMSE of each model decreases as the number of replicas increases, it does so faster in the  Splines model.   

On the other hand, Table \ref{tablebeta23}  shows the coverage obtained for  the period effect ($\beta_2$ and $\beta_3$) using 95\% confidence intervals. The models without splines have a coverage of zero in all the scenarios because the effect of the period is confused with the time effect in the estimation equation, which is also observed in Figure \ref{beta2RMSE}. The confidence intervals obtained from splines have a coverage close to 95\% when there were 11 experimental units per sequence, see Figure \ref{beta2RMSE}. Results from this simulation suggest that when time effects are not linear or quadratic, models that do not use splines overestimate the variance of the treatment effects,  and these fail to estimate unbiasedly the period effects. This leads to erroneous conclusions about the effectiveness  of the sequences and treatments.

\section{Application}

\begin{figure}[ht]
\centering % centering figure
\includegraphics[width=13cm]{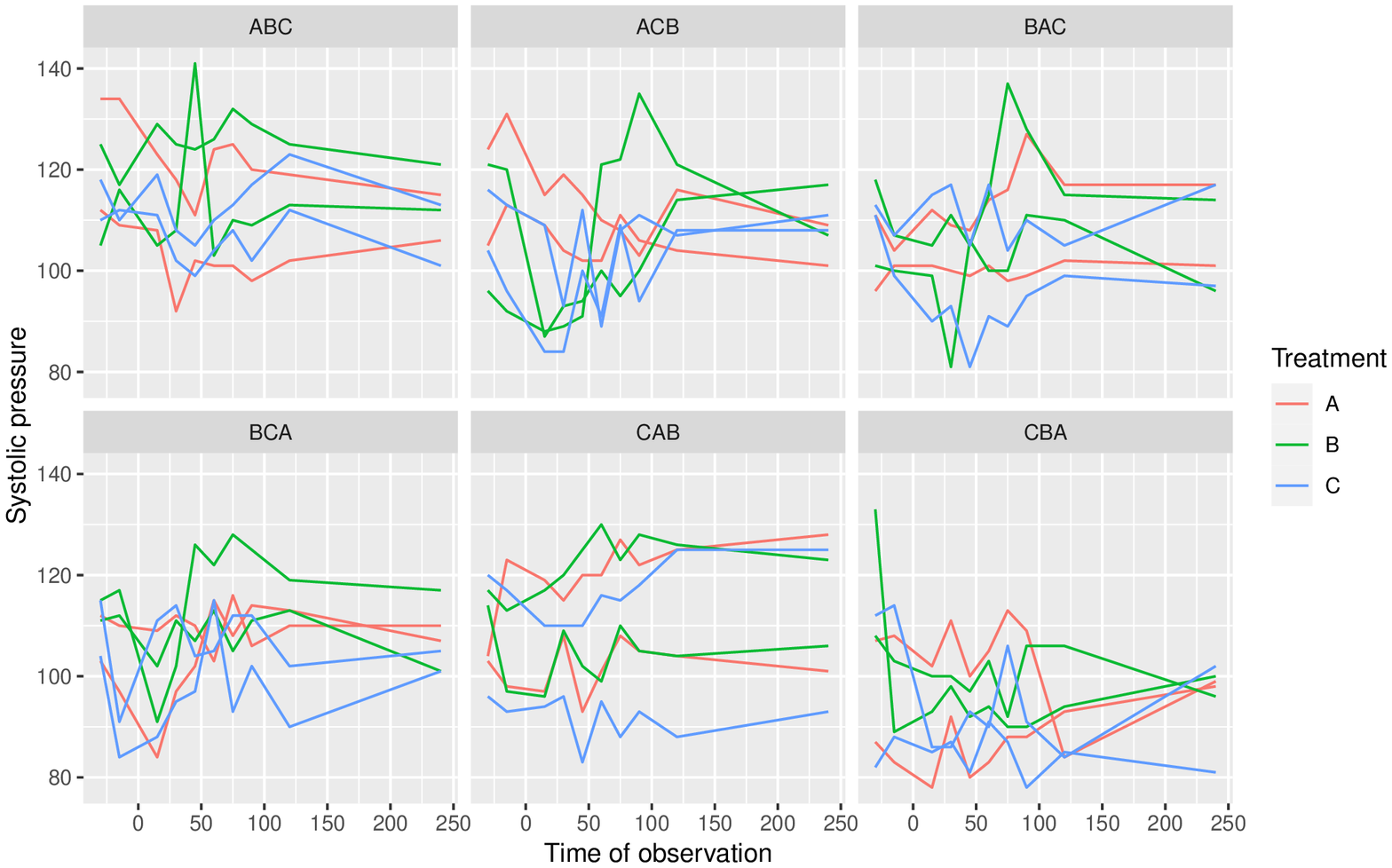}
\caption{Blood systolic pressure (mmHg) observed through time (minutes)}
\label{datosreales}
\end{figure}

\begin{figure}[ht]
\centering % centering figure
\includegraphics[width=13cm]{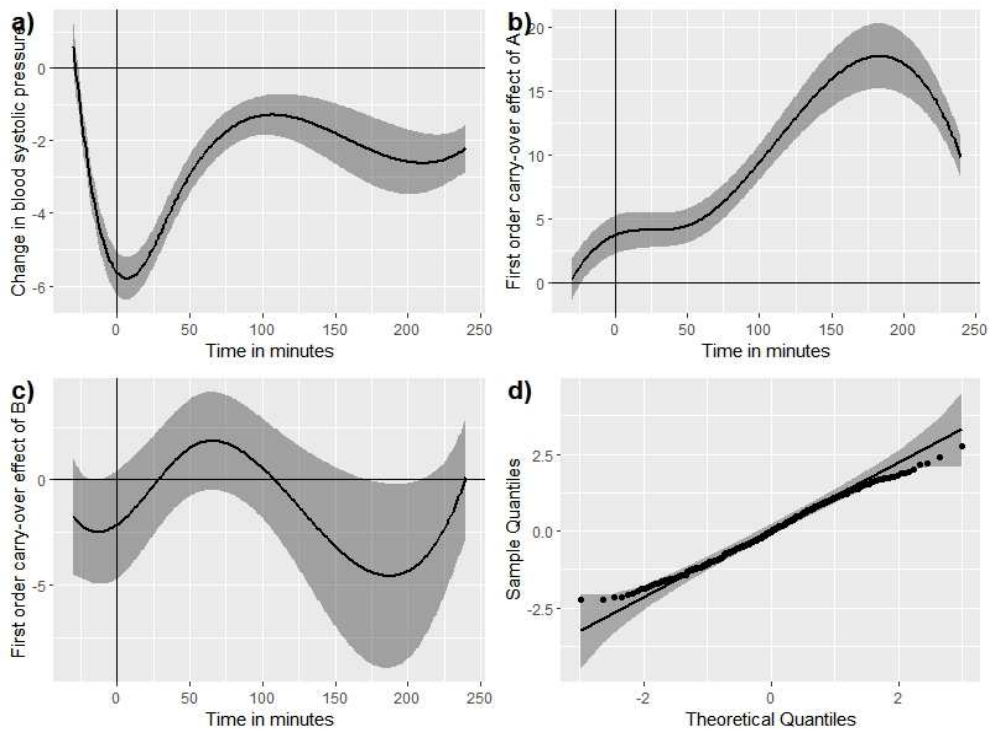}
\caption{a) Changes of blood systolic pressure through time using splines, b) First order carry-over effect of the treatment A  through time using splines, c) First order carry-over effect of the treatment B  through time using splines, and d) Residuals normal probability plot. All figures present 95\% confidence intervals and are based on the cross-over design of Table \ref{tabla100}}
\label{tf_splines}
\end{figure}
Two studies are presented below where the model proposed in \eqref{ecverdad} is used. In both the software \citet{RRR} is used through adaptation of the package \textbf{geeM} built by \citet{geeM}.

\subsection{Systolic pressure data}

\begin{table}[ht]
 \centering    \captionsetup{justification=centering,margin=2cm}
\begin{tabular}{rrrrr}
  \hline
 & Estimate & Std.err & Wald & Pr($>|W|$) \\ 
  \hline
Intercept & 109.35 & 3.17 & 1192.44 & 0.00 \\ 
  BaseLine & 3.85 & 1.47 & 6.82 & 0.01 \\ 
  Period 2 & -0.88 & 3.49 & 0.06 & 0.80 \\ 
  Period 3 & -3.22 & 3.60 & 0.80 & 0.37 \\ 
  Treatment B & 0.70 & 3.04 & 0.05 & 0.82 \\ 
  Treatment C & -5.95 & 3.60 & 2.73 & 0.01 \\ 
   \hline
\end{tabular}
\caption{Analysis of blood systolic pressure data using GEE-Splines}
\label{tableGEEsangre}
\end{table}
\citet[pg 204]{ken15} describes the following crossover design: 3 treatments for blood pressure control are used; treatment A consists of a 20 mg dose of a test drug, treatment B is a 40 mg dose of the same drug, and treatment C is a placebo. 6 sequences of three periods (ABC, ACB, BCA, BAC, CAB, CBA) are organized and each one is applied to two individuals. In each application period, 10 successive measurements of systolic blood pressure are made: 30 and 15 minutes before the application, and 15, 30, 45, 60, 75, 90, 120 and 240 minutes after the application, as shown in Table \ref{tabla100} and the profile is shown in Figure \ref{datosreales}.

Figure \ref{tf_splines}.a) shows the smoothed function corresponding to the effect of time on blood pressure; it is based on the moments of measurement for the design in Table \ref{tabla100}. That is, 30 and 15 minutes before the application, and 15, 30, 45, 60, 75, 90, 120 and 240 minutes after the application. Additionally, Figure \ref{tf_splines}.a) shows the average function and its 95\% confidence bands obtained through cross-validation. A wide drop in pressure is observed from the time that the patient expects to receive treatment, then it rises a little and remains stable. This behavior is widely studied in medical settings, see \cite{stergiou1998white} and \cite{fanelli2021comparison}.

The carry-over effects of treatment A and treatment B are observed in  Figures \ref{tf_splines}.b) and \ref{tf_splines}.c) respectively.
The value for the carry-over effect of A is positive and increases over time, which implies that having applied placebo in a previous period will generate higher blood pressure values in the following application period. 
For treatment B (medium dose), the carry-over effect is close to zero; therefore, it is a negligible effect for the next treatment. 
Table \ref{tableGEEsangre} shows the parametric effects, their standard error and the Wald statistic built from matrices \eqref{an} and \eqref{bn}. It is worth highlighting the positive effect of the baseline, i.e., people have the highest blood pressure before starting the study. The periods are not significant i.e., the conditions were similar across the study, and these had a significant effect on pressure of treatment C on pressure reduction.

Finally, Figure \ref{tf_splines}.d) shows the confidence bands for the quantiles of the standardized residuals defined in \eqref{pearsonres} compared to a standard normal distribution.These residuals seem to fit the normal distribution assumption.
\subsection{Blood Sugar levels in Rabbits}
\begin{figure}[ht]
\centering % centering figure
\includegraphics[width=13cm]{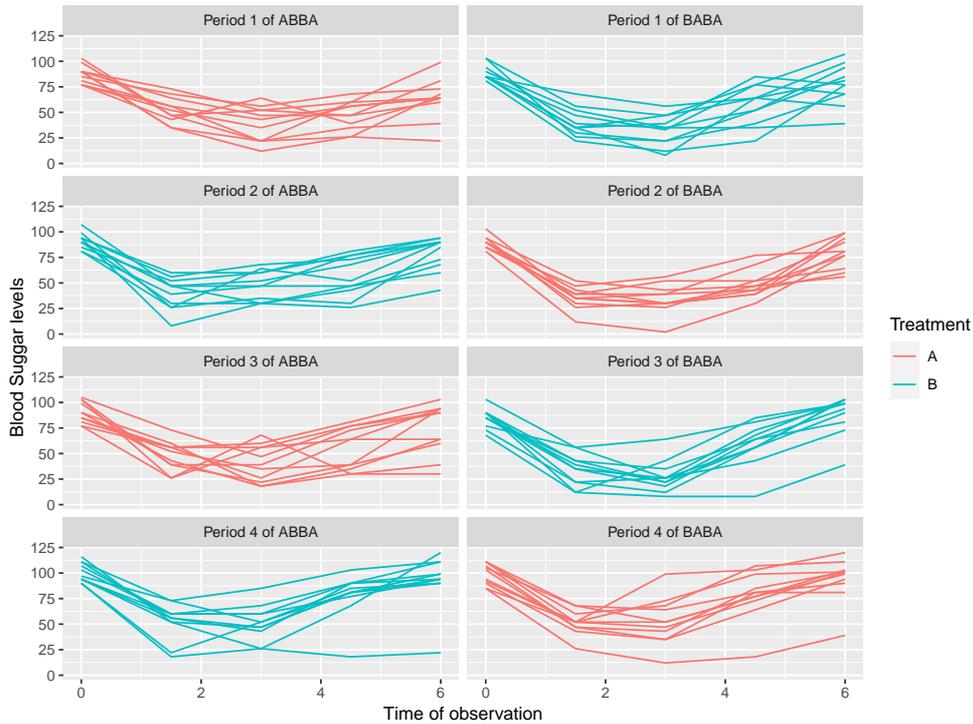}
\caption{Blood sugar levels (mg/dL) in rabbits through time (hours)}
\label{datosreales1}
\end{figure}

\citet{10.1093/biostatistics/kxp046} described the following cross-over experiment: two treatments for the control of diabetes A and B were used, two sequences of four periods (ABAB, BABA) are organized and each one was applied to twelve female rabbits; each period lasted one week.
In each period, at the middle of the week, five successive measurements of the blood sugar level were taken: 0, 1.5, 3, 4.5 and 6 hours after the application, as shown in Table \ref{tabla1001} and  Figure \ref{datosreales1}.

\begin{figure}[ht]
\centering % centering figure
\includegraphics[width=10cm]{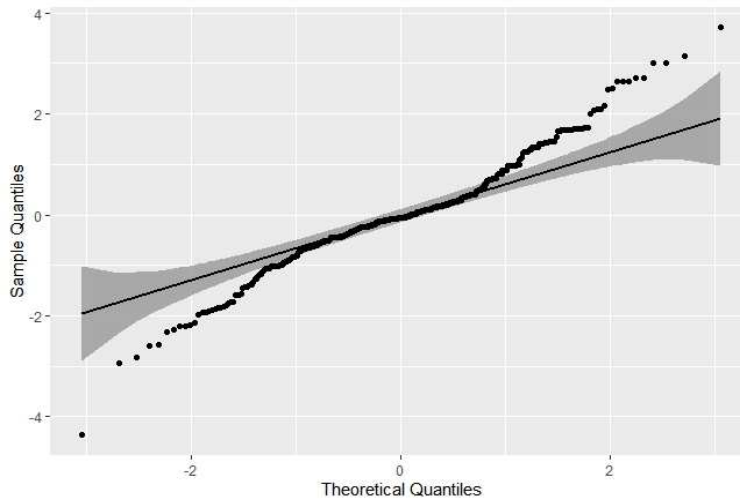}
\caption{Residuals normal probability plot with 95\% confidence intervals of the residuals obtained assuming a Gaussian distribution of the response in the cross-over design of the Table  \ref{tabla1001}}
\label{tf_normal_qq}
\end{figure}

\begin{figure}[ht]
\centering % centering figure
\includegraphics[width=13cm]{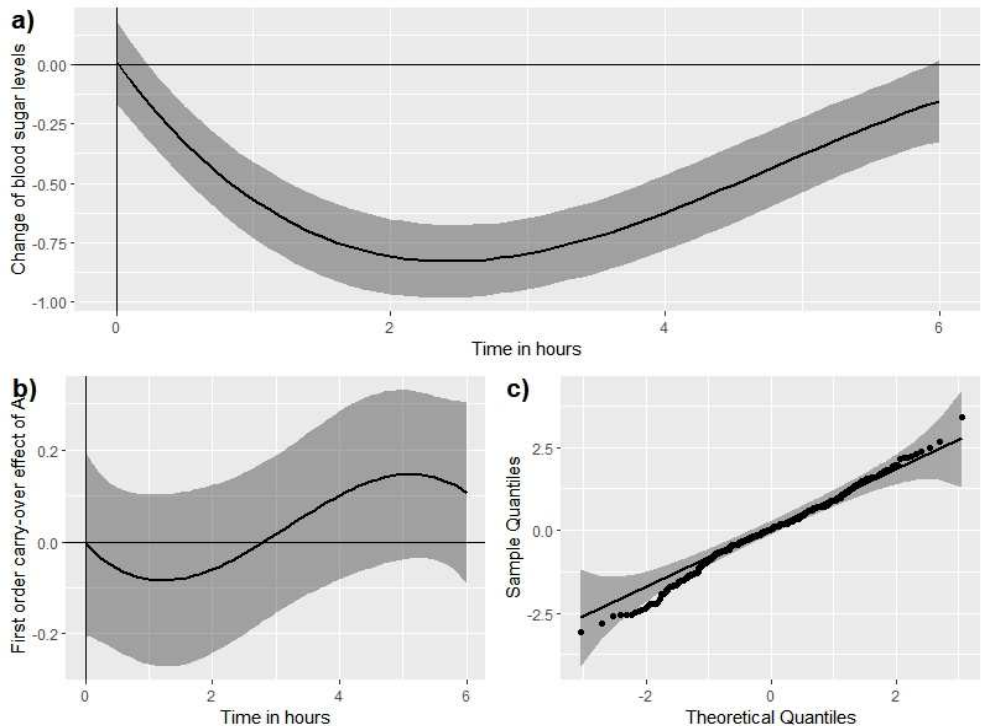}
\caption{a) Change of blood sugar levels through time using splines, b) First order carry-over effect of the treatment A  through time using splines, and c) Residuals normal probability plot. All figures present 95\% confidence intervals and are based on the cross-over design of the Table  \ref{tabla1001} with linear log link for the mean of an assumed gamma distribution}
\label{tf_splines_rabbits}
\end{figure}

\begin{table}[ht]
     \centering    \captionsetup{justification=centering,margin=2cm}
    \captionsetup{justification=centering,margin=2cm}
   \begin{tabular}{c|c|c|}
        Sequence & Period 1 & Period 2\\
        \hline
        \begin{tabular}{cc}
            (1) ABAB & Ind 1\\
            &$\vdots$\\
             & Ind 11
        \end{tabular} & \begin{tabular}{c}
               5 observations\\
                $\vdots$\\
             5 observations
        \end{tabular}&  \begin{tabular}{c}
              5 observations\\
               $\vdots$\\
              5 observations
        \end{tabular} \\
         \begin{tabular}{cc}
            (2) BABA & Ind 12\\
             &$\vdots$\\
             & Ind 22
        \end{tabular} & \begin{tabular}{c}
               5 observations\\
               $\vdots$\\
               5 observations
        \end{tabular}&  \begin{tabular}{c}
              5 observations\\
               $\vdots$\\
               5 observations
        \end{tabular}\\
        \hline
    \end{tabular}
    \caption{Structure of the cross-over design of blood sugar levels in rabbits}
    \label{tabla1001}
\end{table}
%Partiendo del supuesto de que la distribución de los niveles de azúcar en sangre es normal,  se ejecutó un primer análisis. Al realizar la gráfica de probabilidad normal de los residuos estandarizados definidos en \eqref{pearsonres}, se observa que no se ajustan correctamente a una distribución normal estándar, como se ve en la Figura \ referencia{tf_normal_qq} y el supuesto es rechazado. Luego se explora una distribución gamma, con enlace loglineal. La Figura \ref{tf_splines_rabbits} C) muestra las bandas de confianza para los cuantiles de los residuales estandarizados definidos en \eqref{pearsonres} frente a una distribución normal estándar, concluyendo que el supuesto de distribución gamma es adecuado. Por lo tanto, todo el análisis se realiza con el supuesto de distribución gamma y el modelo loglineal dado por:
Assuming that the distribution of blood sugar levels is normal, a first analysis was run. When making the normal probability plot of the standardized residuals defined in \eqref{pearsonres}, it is observed that they do not fit correctly to a standard normal distribution, as seen in Figure \ref{tf_normal_qq} and the assumption is rejected. A gamma distribution is then explored, with loglinear linkage. Figure \ref{tf_splines_rabbits} c) shows the confidence bands for the quantiles of the standardized residuals defined in \eqref{pearsonres} against a standard normal distribution, concluding that the gamma distribution assumption is adequate.
%Para  comparar las distribuciones, se realizan tres modelos para analizar la variable respuesta: i) Bajo el supuesto de normalidad, ii) Bajo el supuesto de distribución gamma y enalace inverso y iii) bajo el supuesto de distribución gamma y un enlace loglineal. En cada uno el QIC es calculado y se obtiene la tabla \ref
To compare the distributions, three models are made to analyze the response variable: i) Under the assumption of normality, ii) Under the assumption of gamma distribution and inverse link, and iii) under the assumption of gamma distribution and a loglinear link. In each one, the QIC is calculated and shows in Table \ref{tabla800}. 
\begin{table}[ht]
     \centering    \captionsetup{justification=centering,margin=2cm}
    \captionsetup{justification=centering,margin=2cm}
   \begin{tabular}{|c|c|}
   \hline
Model & QIC \\
\hline
Normal & 3927.6\\
Gamma Inverse & 3732.22\\
Gamma Log & 3728.5 \\
\hline
    \end{tabular}
    \caption{QIC for the three fitted models to the response of blood sugar levels in rabbits}
    \label{tabla800}
\end{table}
Therefore, all the analysis is carried out with the gamma distribution assumption and the loglinear model given by:
\begin{align*}
      \ln(\mu_{ijk})&=\pmb{x}_{ijk}^T\pmb{\beta} + f_1(\pmb{Z}_{1ijk})+ f_2(\pmb{Z}_{2ijk}) 
\end{align*}

The spline-smoothed function for the time effect on the blood sugar level of female rabbits is shown in Figure \ref{tf_splines_rabbits}a), it is based on the moments of measurement for the design of Table \ref{tabla1001}; the average functions and their 95\% confidence bands, estimated through cross validation are presented. A marked decrease in levels is observed until 2 hours, and then an increase until hour 6. In \citet[pag 237]{ken15} it was stated that there was an effect of the hours, but its form is not explained. However, the proposed model permits to describe this effect. Figure \ref{tf_splines_rabbits}.b) shows the carry-over effects of treatment A over treatment B. This increases over time, but it is close to 0, which implies that having applied A in a previous period will not significantly affect the next application period. The parametric effects, their standard error and the Wald statistic constructed with matrices \eqref{an} and \eqref{bn} are presented in Table \ref{tableGEEsangre1}.
\begin{table}[ht]
 \centering    \captionsetup{justification=centering,margin=2cm}
\begin{tabular}{rrrrr}
  \hline
 & Estimate & Std.err & Wald & Pr($>|W|$) \\ 
  \hline
Intercept  & 4.47 & 0.05 & 9734.05 & 0.00 \\ 
  Period 2 & 0.01 & 0.06 & 0.00 & 0.96 \\ 
  Period 3 & -0.01 & 0.06 & 0.01 & 0.94 \\ 
  Period 4 & 0.25 & 0.06 & 16.68 & 0.00 \\ 
  Treatment B& -0.02 & 0.04 & 0.28 & 0.60 \\
   \hline
\end{tabular}
\caption{Analysis of blood sugar levels data using GEE-Splines}
\label{tableGEEsangre1}
\end{table}
It is noteworthy that there are no significant effects of treatment, similar to that obtained by \citet{ken15} and \citet{10.1093/biostatistics/kxp046}, but there is an positive effect of period four.
This behavior was not analyzed in previous studies and can be seen as increased insulin resistance by blood cells; similar to behaviors reported by \citet{ning2015high} and \citet{da2020role}.
%En el modelamiento de la presión arterial la distribución normal presenta un mejor desempeño que la gamma, mientras que en los niveles de azucar en sangre, la distribución gamma logra un mejor ajuste que la distribución normal. Elegir la distribución más adecuada es deseable, porque los errores estandar de cada estimador de los efectos parametricos del modelo son menores. Además los residuales de pearson presentan un comportamiento que se ajusta a una normal estandar en ambos casos, garantizando que la especificación del modelo es adecuada.

In the modeling of blood pressure, the normal distribution presents a better performance than the gamma distribution, while in blood sugar levels, the gamma distribution achieves a better fit than the normal distribution. Choosing the most suitable distribution is desirable, because the standard errors of each estimator of the parametric effects of the model are smaller. In addition, the Pearson residuals show a behavior that conforms to a standard normal in both cases, guaranteeing that the model specification is adequate.

These applications show the importance of the semiparametric approach proposed in this paper to model time and carry-over effects in crossover designs with repeated measures within periods. Also, it complements the simulation study, where the efficiency of the proposed model over conventional models was verified.

\section{Conclusions}

The proposed methodology provides highly desirable properties of the resulting estimators. It allows doing asymptotic inference and better to model temporal carry-over behaviors that would be intractable in parametric scenarios, as in the case of blood pressure data, where these effects do not present the typical polynomial effects. In addition, detecting these carry-over effects of the placebo allows estimating treatment effects with greater precision and unbiasedness, which is basically the objective of any crossover design.
\\
In the insulin data in rabbits, the behavior of the estimated effect of time is similar to a quadratic form, which shows that this methodological proposal encompasses the classical parametric temporal models with linear or cubic polynomials. In addition, the GEE allow modeling a large number of response variables, not only normal or continuous, but also counts or proportions of successes.
In the simulation, the inferential gain is evidenced in terms of coverage and control of the type I and  II error of the hypothesis tests associated with the parameters of interest; that is, treatment and period effects when the temporal behavior is sinusoidal.
While linear or quadratic models lose efficiency and unbiasedness; therefore, estimation with splines is presented as a useful tool for this type of design.
\\
The asymptotic properties of the estimators allow an agile and fast verification of the model, because its similarity with weighted least squares is demonstrated.
Therefore, the adaptation of widely used diagnostic tests in normal linear models can be used. 
%In the application exercises, the fit of the residuals to the confidence bands of a standard normal is displayed.

% Futuros trabajos, bayesiano o funcionales

\appendix
\section{Appendix 1}\label{Apen1}

\begin{proof}
 Let
\begin{equation}
    \pmb{\theta}(\pmb{\hat{\beta}}, \pmb{\hat{\alpha}}_1, \pmb{\hat{\alpha}}_2)=\begin{pmatrix}
    \pmb{B}_n^{\frac{1}{2}}(\pmb{\hat{\beta}}-\pmb{\beta})\\
   \sqrt{n} \pmb{H}_{1n}(\pmb{\hat{\alpha}}_1-\pmb{\alpha}_1) + \sqrt{n}\pmb{H}_{1n}\pmb{W}_1^T \pmb{X}(\pmb{\hat{\beta}}-\pmb{\beta})\\
    \sqrt{n} \pmb{H}_{2n}(\pmb{\hat{\alpha}}_2-\pmb{\alpha}_2) + \sqrt{n}\pmb{H}_{2n}\pmb{W}_2^T \pmb{X}(\pmb{\hat{\beta}}-\pmb{\beta})\\
    \end{pmatrix}
\end{equation}
with $\pmb{B}$ as in  Equation \eqref{bn}, and
\begin{align}
    \pmb{W}_{1}&=\left(\pmb{\pi}_{11}, \ldots, \pmb{\pi}_{1n}\right), \;
    \pmb{\pi}_{1i}=(\pi(\pmb{Z}_{1i11}), \ldots, \pi(\pmb{Z}_{1ijn_{ij}}))\nonumber\\
    \pmb{W}_{2}&=\left(\pmb{\pi}_{21}, \ldots, \pmb{\pi}_{2n}\right), \;
    \pmb{\pi}_{2i}=(\pi(\pmb{Z}_{2i11}), \ldots, \pi(\pmb{Z}_{2ijn_{ij}}))\nonumber\\
    \pmb{H}_{1n}&=n\pmb{W}_1^T \pmb{W}_1, \;
    \pmb{H}_{2n}=n\pmb{W}_2^T \pmb{W}_2\label{echn12} \\
    \pmb{X}&=(\pmb{x}_1^T, \ldots, \pmb{x}_n^T), \; \pi(t)=\{s_1(t), \ldots, s_m(t)\}\nonumber
\end{align}
Following ideas presented in \cite{speckman1988kernel} to guarantee that both $\pmb{X}$ and $\pmb{Z}_{ij}$ have finite second moments, we assume that there exists a random variables $\delta_{ijk}$, with $E(\delta_{ijk})=0$ and $Var(\delta_{ijk})\leq\infty$ and continuous functions $g_1 \ldots g_m$  such that:
\begin{equation}
    x_{ijkl}=g_l(\pmb{Z}_{ijk})+ \delta_{ijkl} \qquad 1\leq i \leq n, \; 1\leq j \leq P, \; 1\leq k \leq n_{ij}, \; 1\leq l \leq dim\left(\pmb{\beta}\right) \label{ecdelta}
\end{equation}
These functions allow modeling the possible relationship between the vector of variables associated to the parametric effects and the measurement times within each period.
Let $\pmb{X}_{ij}=(x_{ij1}, \ldots, x_{ijn_{ij}})$ be the parametric effects design matrix, then the following properties hold:
\begin{itemize}
    \item[i)] The succession $\{n_{ij}\}$ is bounded for all $1\leq i \leq n$ and $1\leq j \leq P$, that is:
    \[max(n_{ij})<\infty\]
    \item[ii)] Since $Y_{ijk}$ is a random variable that belongs to  the exponential family and  due to the definition of the generalized estimation equations in \cite{liang1986longitudinal}, and by Lemmma 5.3 given in \citet[pag 116]{lehmann2006theory}, then
    \begin{align}
        E(\pmb{u}_{1i}) &=E\left(\pmb{y}_i-\pmb{\mu}_i\left[\pmb{X}_i \pmb{\beta},\sum_{b=1}^m \alpha_{1b}s_b(t)  ,\hat{f}_2(\pmb{Z}_{2i}) \right]\right)=0\nonumber
          \end{align} 
Therefore, the expected value of \eqref{u1} is:
 \begin{align}
         E(U_1(\pmb{\epsilon}_i, t))&=E\left\{\sum_{i=1}^n \left\{ diag\left(\frac{\partial \mu_{ijk}}{\partial \pmb{\alpha}_1}\right)\right\}_i \frac {\pmb{V}^{-1}_{1i}}{\phi} \left(\pmb{y}_i-\pmb{\mu}_i\left[\pmb{X}_i \pmb{\beta},\sum_{b=1}^m \alpha_{1b}s_b(t)  ,\hat{f}_2(\pmb{Z}_{2i}) \right]\right)\right\} \nonumber\\
         &=\sum_{i=1}^n \left\{ diag\left(\frac{\partial \mu_{ijk}}{\partial \pmb{\alpha}_1}\right)\right\}_i \frac {\pmb{V}^{-1}_{1i}}{\phi} E \left(\pmb{y}_i-\pmb{\mu}_i\left[\pmb{X}_i \pmb{\beta},\sum_{b=1}^m \alpha_{1b}s_b(t)  ,\hat{f}_2(\pmb{Z}_{2i}) \right]\right)\nonumber\\
         &=0 \, \forall t\in \mathbb{R}\nonumber
    \end{align}
Analogous results are obtained for equations \eqref{u2}, \eqref{u3} and \eqref{u4}. As $E(Y_{ijk}^2)< \infty$  and  the density function satisfies the regularity conditions then,  by Theorems 1 and 2 of \cite{panv} and by theorem 2.6 of \citet[pg 440]{lehmann2006theory}, for Equation \eqref{u1}, it follows that:
\begin{align}
    0< E\left(U_1(\pmb{\epsilon}_i, t)U_1(\pmb{\epsilon}_i, t)^T\right) <\infty\nonumber
\end{align}
Similarly, for equations \eqref{u2}, \eqref{u3} and \eqref{u4}, the following results are obtained:
    \begin{align*}
        E(U_2(\pmb{\epsilon}_i, t))&=0, \qquad 0< E\left(U_2(\pmb{\epsilon}_i, t)U_2(\pmb{\epsilon}_i, t)^T\right) <\infty, \; \forall t\in \mathbb{R},\\
        E(U_3(\pmb{\epsilon}_i, t))&=0, \qquad 0< E\left(U_3(\pmb{\epsilon}_i, t)U_3(\pmb{\epsilon}_i, t)^T\right) <\infty, \; \forall t\in \mathbb{R},\\
        E(U_4(\pmb{\epsilon}_i))&=0, \qquad 0< E\left(U_4(\pmb{\epsilon}_i)U_4(\pmb{\epsilon}_i)^T\right) <\infty,
    \end{align*}
where $\pmb{\epsilon}_i=(\epsilon_{i11}, \ldots, \epsilon_{in_{iP}})=\pmb{z}_i-\pmb{\hat{z}}_i$, with  $\pmb{z}_i$ defined  the equation \eqref{eczi}.
\item[iii)]According to theorem 2.6 of \citet[pg 441]{lehmann2006theory}, there exist $\{b_{1ijk}\}$, $\{b_{2ijk}\}$ and $\{b_{3ijk}\}$ with $0< inf_{ijk}(b_{ijk})\leq sup_{ijk}(b_{ijk})<\infty$, $0< inf_{ijk}(b_{ijk})\leq sup_{ijk}(b_{ijk})<\infty$ and $0< inf_{ijk}(b_{ijk})\leq sup_{ijk}(b_{ijk})<\infty$ such that when $s\to 0 $, the following properties hold, respectively::
\begin{align}
    \sup_{ijk}\left| E(U_1(\pmb{\epsilon}_{ijk}+s, t))-b_{1ijk}s\right|&=O(s^2), \; \forall t\in \mathbb{R},\nonumber \\
     \sup_{ijk}\left| E(U_2(\pmb{\epsilon}_{ijk}+s, t))-b_{2ijk}s\right|&=O(s^2), \; \forall t\in \mathbb{R},\nonumber \\
     \sup_{ijk}\left| E(U_3(\pmb{\epsilon}_{ijk}+s))-b_{3ijk}s\right|&=O(s^2)\label{ecbetan},
\end{align}
Also, when $s\to 0$, exist constants $c>0$ and,  $C<\infty$ such that:
\begin{align}
    & \sup_{ijk}\left\{ E(U_1(\pmb{\epsilon}_{ijk}+s, t)-U_1(\pmb{\epsilon}_{ijk}, t))^2\right\} \leq C|s|, \; \forall t\in \mathbb{R}\nonumber\\
     &\sup_{ijk}\left\{ E(U_2(\pmb{\epsilon}_{ijk}+s, t)-U_2(\pmb{\epsilon}_{ijk}, t))^2\right\} \leq C|s|, \; \forall t\in \mathbb{R}\nonumber\\
     &\sup_{ijk}\left\{ E(U_3(\pmb{\epsilon}_{ijk}+s)-U_1(\pmb{\epsilon}_{ijk}))^2\right\} \leq C|s|\nonumber
\end{align}
Furthermore, $|U_1(\nu+\eta, t)-U_1(\nu, t)- U_1(\eta, t)|\leq c$, $|U_2(\nu+\eta, t)-U_2(\nu, t)- U_2(\eta, t)|\leq c$ and $|U_3(\nu+\eta)-U_3(\nu)- U_3(\eta)|\leq c$ for any $|\eta|\leq s$ and $ \nu, t \in \mathbb{R}.$
\item Let $\pmb{\Delta}_n$ be a diagonal matrix with elements $\delta_{ijkl}$ defined in  equation \eqref{ecdelta} and let $\pmb{\Lambda}_n$ be a diagonal matrix with elements $b_{3ijk}$ defined in   equation \eqref{ecbetan}, then by definition of the random variables $\delta_{ijkl}$ it follows that:
\begin{align}
    E(\pmb{\Delta}_n)=0 \text{ y } \sup_{n}\left\{ \frac{1}{n}E(||\pmb{\Delta}_n||^2)\right\}<\infty\nonumber
\end{align}
Since, $\pmb{\Gamma}_n $ is a block diagonal matrix with elements $A_i=E\left(U_3(\pmb{\epsilon}_i)U_3(\pmb{\epsilon}_i)^T\right)$, then by the theorem 1 in \cite{panv},
\begin{align}
    &\frac{1}{n}\pmb{\Delta}_n^T \pmb{\Gamma}_n \pmb{\Delta}_n \overset{p}{\to} \pmb{B} \label{ecBs} \\
    &\frac{1}{n}\pmb{\Delta}_n^T \pmb{\Lambda}_n \pmb{\Delta}_n \overset{p}{\to} \pmb{A}\label{ecAs}
\end{align}
\item The matrices $\pmb{H}_{1n}$ y $\pmb{H}_{2n}$ defined in
\eqref{echn12} are symmetric and positive definite, so they have square root \citep{bunch1974triangular}. Let $\pmb{H}_{1n}^{\frac{1}{2}}$ and $\pmb{H}_{2n}^{\frac{1}{2}}$ be those square roots, respectively. Also, as a kernel spline  forms a linearly independent basis, and by Theorem 21.5.1 of \citet[pg 537]{harville}, $\pmb{H}_{1n}^{\frac{1}{2}}$ and $\pmb{H}_{2n}^{\frac{1}{2}}$ are non-singular matrices and their eigenvalues are bounded between zero and infinity.
\end{itemize}
With the previous results and using Theorem 1 and Theorem 2 of \cite{he2002estimation}, the following results are obtained:
\begin{align}
   & \frac{1}{n}\sum_{i=1}^n \sum_{j=1}^{n_i} \left\{\sum_{b=1}^m \hat{\alpha}_{1b}s_b(\pmb{Z}_{1ijk})- f_1(\pmb{Z}_{1ijk})\right\}^2 = O\left(n^{-\frac{2r}{(2r+1)}}\right)\\
   & \frac{1}{n}\sum_{i=1}^n \sum_{j=1}^{n_i} \left\{\sum_{b=1}^m \hat{\alpha}_{1b}s_b(\pmb{Z}_{2ijk})- f_2(\pmb{Z}_{2ijk})\right\}^2 = O\left(n^{-\frac{2r}{(2r+1)}}\right)\\
  &  \sqrt{n}(\hat{\pmb{\beta}}-\pmb{\beta}) \rightarrow N(0, \pmb{A}^{-1}\pmb{B}\pmb{A}^{-1})
 \end{align}
where the matrices $\pmb{A}$ and $\pmb{B}$ are defined in Equations \eqref{an} and \eqref{bn}, respectively.
\end{proof}
\section{Appendix 2}
\begin{table}[ht]
 \centering    \captionsetup{justification=centering,margin=2cm}
\begin{tabular}{|c|c|c|c|}
\hline
 & $H_0:\;\beta_2=3$ &$H_0:\;\beta_3=3$\\  
\hline
\begin{tabular}{c}
n\\
\hline
2\\
     5\\
     8\\
     11\\
     14\\
     17\\
     20\\
     23\\
     26\\
     29\\
     32\\
     35\\
     38\\
     41\\
     44\\
     47\\
     50
\end{tabular} &
\begin{tabular}{ccc}
  \hline
 GEE-S & GEE-1 & GEE2\\ 
  \hline
0.79 & 0.44 & 0.44 \\ 
  0.84 & 0.00 & 0.00 \\ 
  0.94 & 0.00 & 0.00 \\ 
  0.94 & 0.00 & 0.00 \\ 
  0.92 & 0.00 & 0.00 \\ 
  0.94 & 0.00 & 0.00 \\ 
  0.94 & 0.00 & 0.00 \\ 
  0.94 & 0.00 & 0.00 \\ 
  0.94 & 0.00 & 0.00 \\ 
  0.92 & 0.00 & 0.00 \\ 
  0.95 & 0.00 & 0.00 \\ 
  0.92 & 0.00 & 0.00 \\ 
  0.92 & 0.00 & 0.00 \\ 
  0.95 & 0.00 & 0.00 \\ 
  0.93 & 0.00 & 0.00 \\ 
  0.93 & 0.00 & 0.00 \\ 
  0.94 & 0.00 & 0.00 \\ 
   \hline
\end{tabular}
& 
\begin{tabular}{ccc}
  \hline
 GEE-S & GEE-1 & GEE2\\ 
  \hline
0.80 & 0.50 & 0.51 \\ 
  0.89 & 0.00 & 0.00 \\ 
  0.90 & 0.00 & 0.00 \\ 
  0.92 & 0.00 & 0.00 \\ 
  0.92 & 0.00 & 0.00 \\ 
  0.94 & 0.00 & 0.00 \\ 
  0.95 & 0.00 & 0.00 \\ 
  0.95 & 0.00 & 0.00 \\ 
  0.95 & 0.00 & 0.00 \\ 
  0.94 & 0.00 & 0.00 \\ 
  0.95 & 0.00 & 0.00 \\ 
  0.92 & 0.00 & 0.00 \\ 
  0.92 & 0.00 & 0.00 \\ 
  0.92 & 0.00 & 0.00 \\ 
  0.94 & 0.00 & 0.00 \\ 
  0.92 & 0.00 & 0.00 \\ 
  0.90 & 0.00 & 0.00 \\ 
\end{tabular}
\\
\hline
\end{tabular}
\caption{Proportion of times that hypothesis $H_0$ is not rejected for some values of $\beta_2$  and $\beta_3$ (the period effects)}
\label{tablebeta23}
\end{table}
\begin{table}[ht]
 \centering    \captionsetup{justification=centering,margin=2cm}
\begin{tabular}{|c|c|c|c|}
\hline
 & $H_0:\;\beta_1=0.5$ &$H_0:\;\beta_1=1$ & $H_0:\;\beta_1=2$\\  
\hline
\begin{tabular}{c}
n\\
\hline
2\\
     5\\
     8\\
     11\\
     14\\
     17\\
     20\\
     23\\
     26\\
     29\\
     32\\
     35\\
     38\\
     41\\
     44\\
     47\\
     50
\end{tabular} &
\begin{tabular}{ccc}
  \hline
 GEE-S & GEE-1 & GEE2\\ 
  \hline
0.88 & 1.00 & 1.00 \\ 
  0.90 & 1.00 & 1.00 \\ 
  0.93 & 1.00 & 1.00 \\ 
  0.92 & 1.00 & 1.00 \\ 
  0.95 & 1.00 & 1.00 \\ 
  0.94 & 1.00 & 1.00 \\ 
  0.95 & 1.00 & 1.00 \\ 
  0.95 & 1.00 & 1.00 \\ 
  0.96 & 1.00 & 1.00 \\ 
  0.96 & 1.00 & 1.00 \\ 
  0.95 & 1.00 & 1.00 \\ 
  0.95 & 1.00 & 1.00 \\ 
  0.96 & 1.00 & 1.00 \\ 
  0.94 & 1.00 & 1.00 \\ 
  0.95 & 1.00 & 1.00 \\ 
  0.97 & 1.00 & 1.00 \\ 
  0.95 & 1.00 & 1.00 \\ 
   \hline
\end{tabular}
& 
\begin{tabular}{ccc}
  \hline
 GEE-S & GEE-1 & GEE2\\ 
  \hline
0.82 & 1.00 & 1.00 \\ 
  0.92 & 1.00 & 1.00 \\ 
  0.96 & 1.00 & 1.00 \\ 
  0.96 & 1.00 & 1.00 \\ 
  0.94 & 1.00 & 1.00 \\ 
  0.94 & 1.00 & 1.00 \\ 
  0.94 & 1.00 & 1.00 \\ 
  0.96 & 1.00 & 1.00 \\ 
  0.96 & 1.00 & 1.00 \\ 
  0.97 & 1.00 & 1.00 \\ 
  0.94 & 1.00 & 1.00 \\ 
  0.94 & 1.00 & 1.00 \\ 
  0.93 & 1.00 & 1.00 \\ 
  0.94 & 1.00 & 1.00 \\ 
  0.94 & 1.00 & 1.00 \\ 
  0.95 & 1.00 & 1.00 \\ 
  0.94 & 1.00 & 1.00 \\ 
\end{tabular}
& 
\begin{tabular}{ccc}
  \hline
 GEE-S & GEE-1 & GEE2\\ 
  \hline
0.85 & 1.00 & 1.00 \\ 
  0.88 & 1.00 & 1.00 \\ 
  0.93 & 1.00 & 1.00 \\ 
  0.92 & 1.00 & 1.00 \\ 
  0.96 & 1.00 & 1.00 \\ 
  0.94 & 1.00 & 1.00 \\ 
  0.94 & 1.00 & 1.00 \\ 
  0.96 & 1.00 & 1.00 \\ 
  0.97 & 1.00 & 1.00 \\ 
  0.93 & 1.00 & 1.00 \\ 
  0.97 & 1.00 & 1.00 \\ 
  0.92 & 1.00 & 1.00 \\ 
  0.94 & 1.00 & 1.00 \\ 
  0.98 & 1.00 & 1.00 \\ 
  0.93 & 1.00 & 1.00 \\ 
  0.93 & 1.00 & 1.00 \\ 
  0.97 & 1.00 & 1.00 \\

\end{tabular}
\\
\hline
\end{tabular}
\caption{Proportion of times that hypothesis $H_0$ is not rejected for some values of $\beta_1$ (the treatment effect)}
\label{tablebeta}
\end{table}

\end{document}